\documentclass{PoS}
\title{Delta I=3/2 K to pi-pi decays with nearly physical kinematics}

\ShortTitle{Delta I=3/2 K to pi-pi decays with nearly physical kinematics}

\author{\speaker{Elaine Goode}\\
         University of Southampton, School of Physics and Astronomy, Highfield, Southampton, SO17 1BJ, United Kingdom \\
        E-mail: \email{ejg4g08@soton.ac.uk}}

\author{Matthew Lightman\\
        Department of Physics, Columbia University, New York, NY 10027 USA\\
	Department of Physics, Washington University, St. Louis, MO 63130 USA\\
        E-mail: \email{mlightman@physics.wustl.edu}}

\abstract{The $\Delta I = 3/2$ $K \to \pi\pi$ decay amplitude is calculated on RBC/UKQCD $32^3\times 64 $, $L_s=32$ dynamical lattices
with $2+1$ flavors of domain wall fermions using the Dislocation Suppressing Determinant Ratio and Iwasaki gauge action. The calculation is performed close to the 
physical pion mass ($m_{\pi} = 142.9(1.1)$~MeV) and with a single lattice spacing ($a^{-1}= 1.375(9)$~GeV.) We find 
$\text{Re}(A_2)  = (1.436 \pm 0.063_{\rm{stat}} \pm 0.258_{\rm{syst}})\times 10^{-8}\rm{ GeV}$ and 
$\text{Im}(A_2)  = (-6.29 \pm 0.46_{ \rm{stat}} \pm 1.20_{ \rm{syst}})\times 10^{-13}\rm{ GeV}$. These results are combined with
the experimental result for $\epsilon'/\epsilon$ 
to predict $\rm{Im}(A_0) =  -5.32(64)_{\rm{stat}}(71)_{\rm{syst}}\times 10^{-11}\rm{ GeV}$
within the Standard Model. We also perform a reweighting analysis to invesitgate the effects of partial quenching in the light-quark sector of our calculation.
Following reweighting we find $\rm{Re}(A_2)  = (1.52\pm 0.14_{\rm{stat}})\times 10^{-8}\rm{ GeV}$ and 
$\rm{Im}(A_2)  = (-6.47 \pm 0.55_{ \rm{stat}})\times 10^{-13}\rm{ GeV} $, which are consistent with our main
results.}

\FullConference{XXIX International Symposium on Lattice Field Theory \\
                 July 10 -- 16 2011\\
                 Squaw Valley, Lake Tahoe, California}

\usepackage{subfigure}
\usepackage{amsmath}
\usepackage{slashed}

\begin{document}

\section{Introduction}
\vspace{-0.05 in}
The calculation of $K\to\pi\pi$ decay amplitudes is motivated by a desire to understand the $\Delta I = 1/2$ rule
and CP violation in kaon decays. Such a calculation is a non-perturbative problem requiring lattice techniques to make progress.
Previous lattice calculations have relied on the quenched approximation and uncertain chiral extrapolations \cite{Blum:2001xb, kim_christ, CPPACS, chiral_extrap}.
 In this talk we present the results 
from the first realistic lattice calculation of a $K\to\pi\pi$ decay amplitude, where we simulate the two-body decay directly
on the lattice at nearly-physical kinematics.

We proceed by evaluating matrix elements of the $\Delta S = 1$ effective Hamiltonian
\begin{equation}
 \label{eq:Heff}
H_{\text{eff}} = V^*_{us}V_{ud} \sum_i C_i Q_i 
\end{equation}
where $C_i$ are Wilson coefficients and $Q_i$ are four-quark operators. In this talk we consider only the $\Delta I = 3/2$
transition, in which case only three operators contribute in equation 
\eqref{eq:Heff}.
We find it convenient to evaluate unphysical $K^+ \to \pi^+ \pi^+$ matrix elements of the following operators:
\begin{equation}
 Q_{(27,1)} = (\overline{s}^i d^i)_L (\overline{u}^j d^j)_L, \quad  Q_{(8,8)} = (\overline{s}^i d^i)_L (\overline{u}^j d^j)_R \; \text{ and }\;
Q_{(8,8)_{\text{mx}}} = (\overline{s}^i d^j)_L (\overline{u}^j d^i)_R,
\end{equation}
where the operators are labelled according to their transformation under $SU(3)_L \times SU(3)_R$, and the labels $i$ and $j$ on the quark
fields label colour.
These are related to the physical $K^+ \to \pi^+ \pi^0$ matrix elements via the Wigner-Eckart theorem.
\vspace{-0.05 in}
\section{Details of the Simulation \label{sec:simulation}}
\vspace{-0.05 in}
The analysis is performed on a single ensemble of $2+1$ flavour domain wall fermions (DWF) with Dislocation Suppressing Determinant Ratio (DSDR)+Iwasaki
 gauge action at $\beta = 1.75$. 
The lattice size is $32^3 \times 64$ and the extent of the fifth dimension is $L_s=32$. The inverse lattice spacing is $a^{-1}= 1.375(9)$~GeV.
 The ensemble is genereated with sea-quark masses $am_l = 0.001$ and 
$am_h=0.045$, corresponding to a unitary pion mass of approximately 170 MeV. We find the residual mass to be $am_{\rm{res}}=0.00184(1)$. 
 The correlation functions are calculated with
valence quark masses of $m_l=0.0001$ and $m_s=0.049$, corresponding to a valence pion mass of $m_{\pi}=142.9(1.1)$~MeV and kaon mass of
$m_K=511.3(3.9)$~MeV. A total of $63$ gauge configurations, each separated by $8$ molecular dynamics time units, are included in the analysis.

Quark propagators with periodic and antiperiodic boundary conditions in the time direction were computed on each configuration with a source at $t=0$. 
They were then combined so as to effectively double the time extent of the lattice. Meson correlation functions formed using the average of the propagators
with periodic and antiperiodic boundary conditions can be interpreted as containing forward propagating mesons originating at time $t=0$,  whereas
those calculated with the antisymmetric combination can be interpreted as containing backward propagating mesons originating from a source at $t= 64$. 
Strange-quark propagators, also with Periodic $+$ Antiperiodic combinations, were generated with sources at $t_K = $ 20, 24, 28, 32, 36, 40 and 44
in order to calculate $K\rightarrow \pi\pi$ correlation functions with kaon sources at these times, while the two-pion sources remained at either $t=0$ or $t=64$. 
Thus we could achieve time separations between the kaon and two pions of 20, 24, 28 and 32 lattice time units in two different ways which increased the statistics. 
These separations were chosen so that the signals from the kaon and two pions did not decay into noise before reaching the four-quark operator $Q_i$.

For physical $K\to\pi\pi$ decays in the CM frame, the final state pions have equal and opposite non-zero momentum. We achieve this for
the $\pi^+ \pi^+$ final state by giving momentum to the d-quark. The u- and s-quark propagators are generated with 
Coulomb gauge fixed wall sources and periodic spatial boundary conditions. Similarly, the d-quark propagators
used in the zero-momentum pion and two-pion correlation functions are computed with Coulomb gauge fixed wall sources and periodic spatial boundary conditions.
However, the d-quark propagators used in the two-pion and $K\to\pi\pi$ correlation functions with non-zero momentum pions
 are generated with antiperiodic spatial boundary conditions and cosine sources. By imposing antiperiodic spatial boundary conditions
the allowed quark momenta are $  p_n = (\pi + 2\pi n)/L$, where $L$ is the spatial extent of the lattice, corresponding to a ground-state momentum of 
$\pi/L$ in the direction in which antiperiodic boundary conditions have been used. In practice we impose antiperiodic boundary conditions in two spatial directions, which
allows us to simulate pions with ground state momentum $\pm \sqrt{2}\pi/L$. This decision is motivated by the expectation that pions with $p= \sqrt{2}\pi/L$ will
correspond to a two-pion final state with energy close to $m_K$. The use of cosine sources in $K^+ \to \pi^+ \pi^+$ decays is described in \cite{Goode:2010lat}.

\vspace{-0.05 in}
\section{Analysis}
\begin{figure}[t]
\centering
\subfigure[$(27,1)$ operator]{\includegraphics*[width=0.31\textwidth]{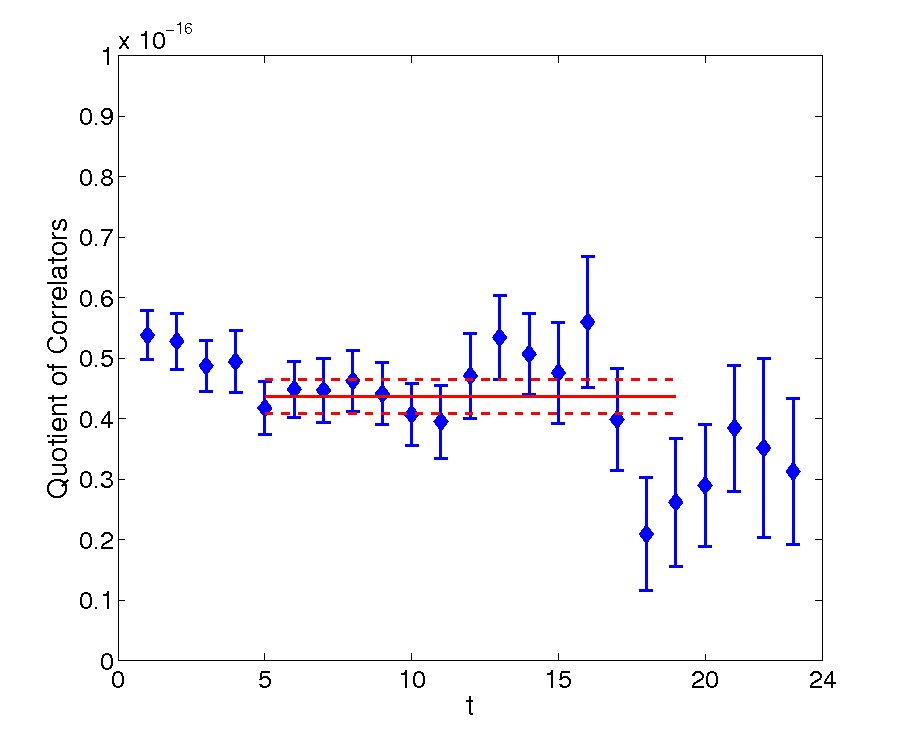}}
\subfigure[$(8,8)$ operator]{\includegraphics*[width=0.31\textwidth]{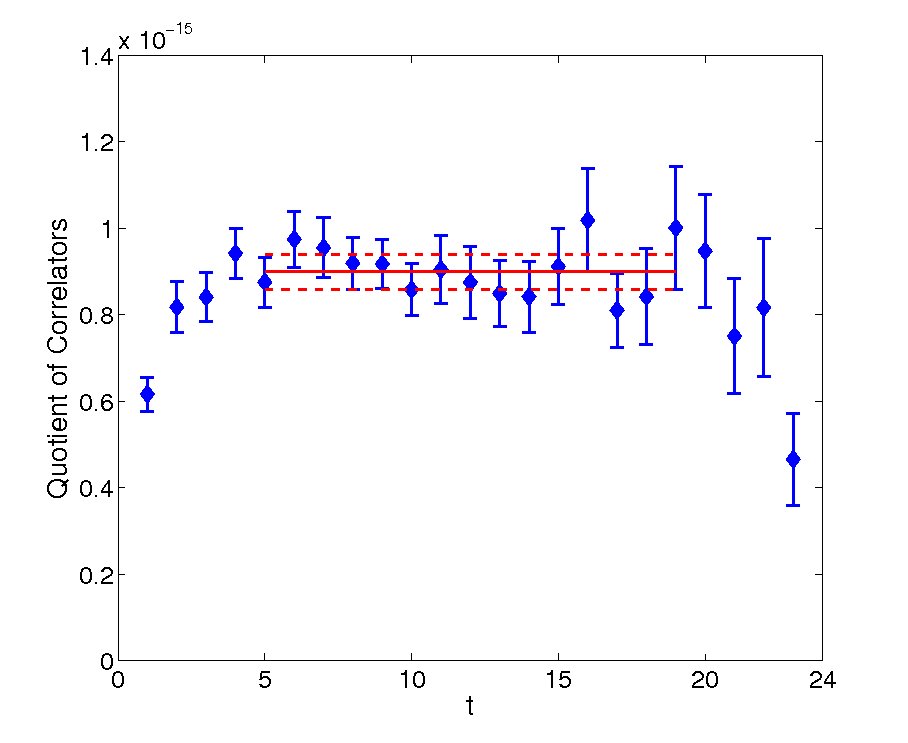}}
\subfigure[$(8,8){\text{mix}}$ operator]{\includegraphics*[width=0.31\textwidth]{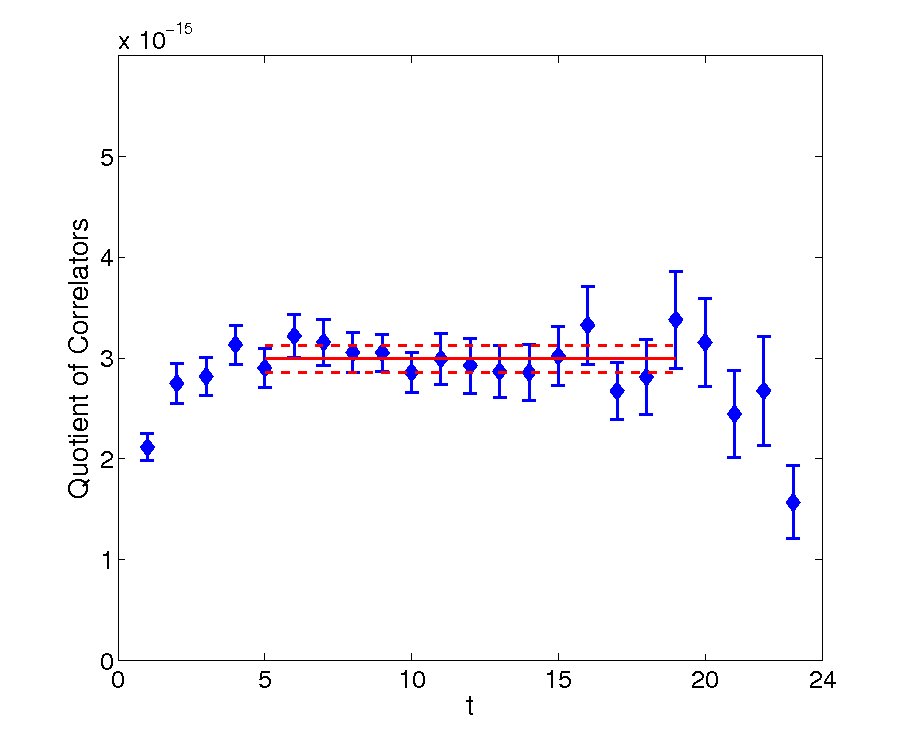}}
\caption{\label{fig:q_plot} $K\rightarrow \pi\pi$ quotient plots for $p= \sqrt{2}\pi/L$.
The two pion source is at $t=0$ while the kaon source is at $t=24$. The dashed line shows the error on the fit.}
\end{figure}

\vspace{-0.05 in}
We extract the $K \rightarrow \pi\pi$ matrix element $\mathcal{M}$ by fitting a constant to the left hand side of \eqref{eq:quot1}

\begin{equation}
 \frac{C^i_{K\pi\pi}(t)}{C_K(t_K-t)C_{\pi\pi}(t)} = \frac{\mathcal{M}_i}{Z_K Z_{\pi\pi}}.
\label{eq:quot1}
\end{equation}
$C^i_{K\pi\pi}$ is the $K \rightarrow \pi\pi$ correlator with a
kaon source at $t_K$, $i$ labels the four-quark operator $Q_i$ which is inserted at time $t$, and $Z_K$ and $Z_{\pi\pi}$ are calculated from the kaon and 
two-pion correlators respectively, whose sources are at $t=0$.
The left hand side of equation \eqref{eq:quot1} is plotted in Figure \ref{fig:q_plot} for each operator. 
The figure demonstrates that sufficiently far from 
the kaon and two-pion sources we are justified in fitting to a constant. The fit results for $\mathcal{M}_i/(Z_K Z_{\pi\pi})$ are indicated on the plot.

We also use a quotient method to extract the two-pion energy, as we find this improves the
statistical precision of the fits. We fit the quotient of correlators 
$ C_{\pi\pi}/(C_{\pi})^2 \sim A e^{-\Delta E\, t}$
to extract $\Delta E= E_{\pi\pi} - 2E_{\pi}$. We then get the two-pion energy by calculating $\Delta E+ 2 E_{\pi}$, where in the case of
$p=0$, $E_{\pi}=m_{\pi}$ while for $p=\sqrt{2}\pi/L$, $E_{\pi}$ is found from a 2 parameter fit to the pion correlation function which also
has $p=\sqrt{2}\pi/L$. The numerical results for all the meson masses and energies which we extract from the correlation functions
are given in Table \ref{tab:masses}. From Table \ref{tab:masses} we see that the kaon mass is not exactly equal to
the two-pion energy, so our calculation is not quite on-shell. This is taken into account when estimating the systematic error on the final results.


\begin{table}[h]
\begin{center}\vspace{0.15in}
\begin{tabular}{|c|c|c|c|c|c|c|}
\hline
units&$m_{\pi}$ & $m_K$ & $E_{\pi,2}$ & $E_{\pi\pi,0}$ & $E_{\pi\pi,2}$ & $m_K - E_{\pi\pi,2}$ \\
\hline
lattice &0.10395(32) & 0.37193(91) & 0.1737(14) & 0.20948(63)& 0.3583(35) & 0.0136(35)\\
MeV &142.9(1.1) & 511.3(3.9) & 238.8(2.4)& 288.0(2.2)& 492.6(5.5) &18.7(4.8)\\
\hline
\end{tabular}
\caption{Results for meson masses and energies. The subscripts $0$, $2$ denote
$p= 0$, $p = \sqrt{2}\pi/L$ respectively. \label{tab:masses}}\end{center}
\end{table}

The finite volume matrix elements are related to the infinite volume amplitudes $\mathcal{A}_i$ (where $i$ labels the 4-quark operator) using the Lellouch-L\"uscher factor
\cite{LLfactor, LLfactor_sachrajda} . 
In particular we have
\begin{equation}
 \mathcal{A}_i = \frac{1}{2} \left[ \frac{1}{\pi q_{\pi}} \sqrt{\frac{\partial \phi}{\partial q_{\pi}} + \frac{\partial \delta}{\partial q_{\pi}} } \right]
  L^{3/2} \sqrt{m_K} E_{\pi\pi} \mathcal{M}_i
\label{eq:amp}
\end{equation}
where the quantity in square brackets 
contains the effects of the Lellouch-L\"uscher factor beyond the free field normalization, $\delta$ is the s-wave phase shift,  
$q_{\pi}$ is a dimensionless quantity related to the individual pion momentum $k_{\pi}$ via
$q_{\pi} = k_{\pi}L/2\pi$ and $\phi$ is a kinematic function defined in \cite{LLfactor}.
The pion momentum $k_{\pi}$ is calculated using the dispersion relation $E_{\pi\pi} = 2\sqrt{k_{\pi}^2 + m^2_{\pi}}$,
and differs from $\sqrt{2}\pi/L$ due to interactions between the two pions.  Once $q_{\pi}$ is known, $\delta$ can be calculated using the
L\"uscher quantisation condition \cite{Luscher_quant}, $ n\pi = \delta(k_{\pi}) + \phi(q_{\pi})$.
As discussed in \cite{Goode:2010lat}, the phase shift derivative is calculated using the phenomenological curve of \cite{Schenk:1991xe}.
This is necessary because we only have two values of the two-pion energy from which to extract the phase shift.

The amplitudes $\mathcal{A}_i$ are related to the physical decay amplitude $A_2$ via
\begin{equation}
 A_2 = a^{-3} \sqrt{\frac{3}{2}} G_F V_{ud} V^*_{us} \sum_{i,j} C_i(\mu) Z_{ij}(\mu) \mathcal{A}_j,
\end{equation}
where $C_i$ are the Wilson coefficients and $Z_{ij}$ are the renormalization constants. The Wilson coefficients must be evaluated
at the same scale and scheme as the renormalization constants. 
The renormalization constants are first evaluated in the RI-SMOM($\slashed{q}, \slashed{q}$) scheme \cite{Garron_PoS}. In order to minimize
discretization effects, this procedure takes place at a relatively low energy $\mu_0 = 1.145$~GeV. A non-perturbative step-scaling 
function is then used to convert these results to a scale of $3$~GeV, at which point a perturbative matching to the $\overline{\text{MS}}$-NDR scheme is
possible. The Wilson coefficients are known in the NDR scheme at the W-mass scale, and can be perturbatively run to the desired matching point of
3~GeV \cite{Buchalla:1995vs}.
\vspace{-0.05 in}
\section{Results}
\vspace{-0.05 in}
We calculate $A_2$ for the four different separations between the kaon source and two-pion source. Our final result, presented in equation
\eqref{eq:A2}, is an error weighted average over these four results. The first error in equation \eqref{eq:A2} is a statistical error,
where the statistical uncertainties in the amplitude $A_i$ and lattice spacing (4\% in total) are combined in quadrature with the statistical
error on the renormalization constants (0.8\% for Re$A_2$ and 6\% for Im($A_2$)). 
\begin{equation}
 \label{eq:A2}
\begin{split}
\text{Re}(A_2) & = (1.436\pm 0.063_{\text{stat}}\pm 0.258_{\text{syst}})\times 10^{-8}\text{ GeV},\\ 
\quad \text{Im}(A_2) &= (-6.29 \pm 0.46_{\text{stat}} \pm 1.20_{\text{syst}})\times 10^{-13}\text{ GeV}.
\end{split}
\end{equation}

The second error in equation \eqref{eq:A2} is systematic. 
The systematic errors in our calculation, for (Re($A_2$), Im($A_2$)) respectively, are from lattice artifacts (15\%, 15\%), 
uncertainty in the phase shift derivative (0.32\% 0.32\%), finite volume (6.2\% and 6.8\%), partial quenching (3.5\% and 1.7\%),
uncertainties in the renormalization procedure (1.7\% and 4.7\%), unphysical kinematics (3.0\% and 0.22\%), and perturbative truncation in the 
evaluation of the Wilson coefficients (7.1\% 8.1\%). Combining in quadrature, we find the systematic errors
to be 18\% for Re($A_2$) and 19\% for Im($A_2$). Further details on how these errors are estimanted can be found in \cite{ktopipi, lightman_thesis}.

\vspace{-0.05 in}
\section{Reweighting}
\begin{figure}[h]
\centering
\subfigure[Reweighting Re(A$_2$)\label{fig:rwReA2}]{\includegraphics[width=0.45\textwidth]{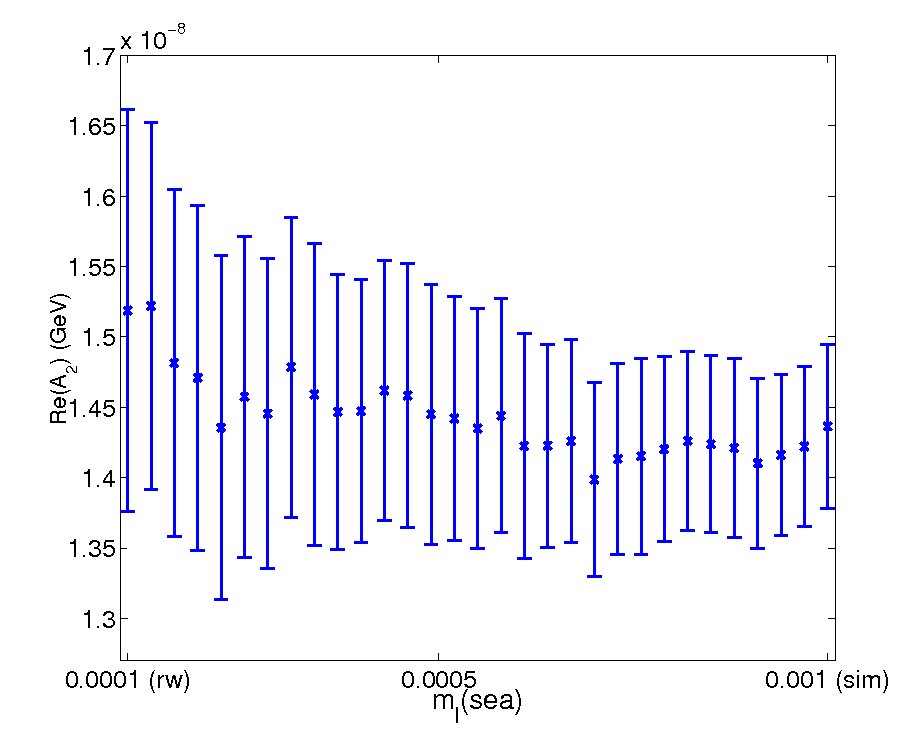}}
\subfigure[Reweighting Im(A$_2$)\label{fig:rwImA2}]{\includegraphics[width=0.45\textwidth]{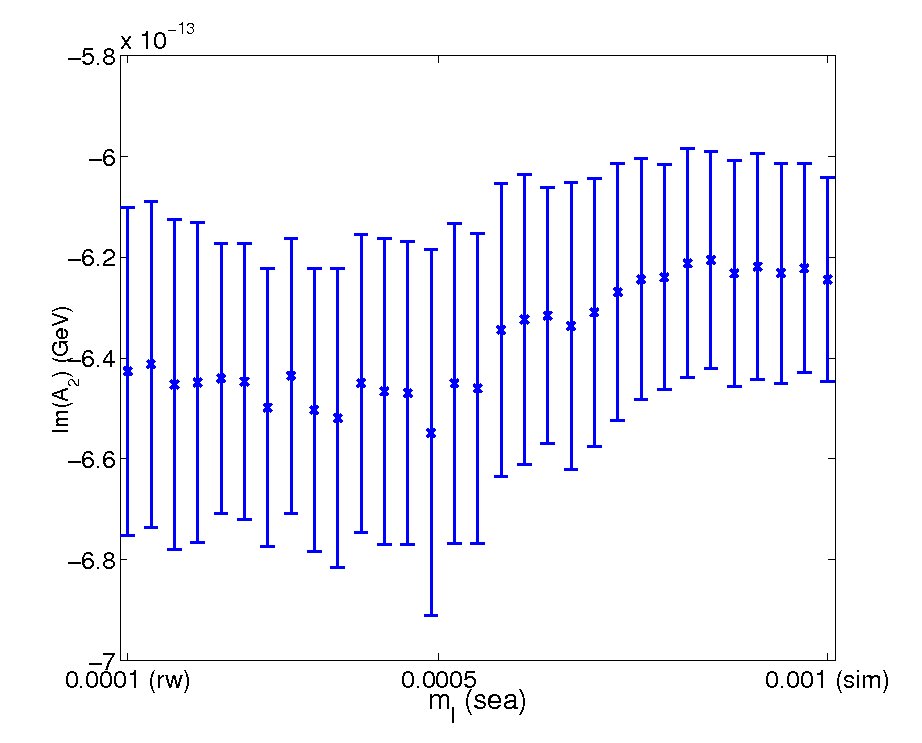}}
\caption{Reweighting $\text{A}_2$ from $m_l^{\mathrm{sea}}=0.001$ to $m_l^{\mathrm{sea}}=0.0001$. 
\label{fig:rw}}
\end{figure}

\vspace{-0.05 in}
We use the technique of reweighting to test the consequences of the partial quenching in the light-quark sector of our calculation.
The reweighting is performed in 30 increments from the simulated mass $m_l^{\text{sea}}=0.001$ down to a value of $m_l^{\text{sea}}=0.0001$, 
corresponding to the valence light-quark mass. The results are shown in Figure~\ref{fig:rw}.
The rightmost point in Figure \ref{fig:rw}(a) shows the result for $\text{Re(A}_2)$ before reweighting, 
while the remaining points show the results after reweighting to the mass indicated on the $x$-axis, 
ending with $m_l^{\mathrm{sea}}=0.0001$ for the leftmost point. 
Similarly Figure \ref{fig:rw}(b) shows the effects of reweighting on $\text{Im(A}_2)$.  Examining the figures, it can be seen that
the errors on $\text{Re}(A_2)$ and $\text{Im}(A_2)$ grow, but the central values remain unchanged within the errors. 
Since reweighting effectively reduces the number of configurations contributing to the observables \cite{Aoki:2010dy} 
it is natural that  the statistical error should increase. However, the observation that the central values are unchanged confirms that 
partial quenching in the light quark does not introduce a significant source of systematic error.

The final results after reweighting are shown in Table \ref{tab:rw_A2} where they are compared with the results before reweighting.

\begin{table}[h]
\centering
\begin{tabular}[h]{|c|c|c|}
\hline
 & $m_l = 0.001$ & $m_l = 0.0001$ (reweighted) \\ \hline
Re(A$_2$) & $1.436(63)\times 10^{-8}$ GeV& $1.52(14)\times 10^{-8}$ GeV \\
Im(A$_2$) & $-6.29(46)\times 10^{-13}$ GeV& $-6.47(55)\times 10^{-13}$ GeV\\
\hline
\end{tabular}
\caption{\label{tab:rw_A2} $A_2$ before and after reweighting.}
\end{table}

\vspace{-0.05 in}
\section{Prediction for Im($A_0$)}
\vspace{-0.05 in}
Assuming isospin symmetry, the CP-violating parameter $\epsilon'/\epsilon $ can be expressed in terms of Re$(A_0)$, Im$(A_0)$, Re$(A_2)$ and Im$(A_2)$
according to equation \eqref{eq:cp}:

\begin{equation}
\label{eq:cp}
 \text{Re}\left ( \frac{\epsilon'}{\epsilon} \right) = \frac{\omega}{\sqrt{2}\left\vert \epsilon \right \vert} \left[ \frac{\text{Im} (A_2)}{\text{Re}(A_2)} - \frac{\text{Im} (A_0)}{\text{Re}(A_0)} \right]\,,
\end{equation}
where $\omega = \text{Re}(A_2)/\text{Re}(A_0)$. 
$\text{Re}\left ( \frac{\epsilon'}{\epsilon} \right)$, $|\epsilon|$, $\omega$ and $\text{Re}(A_0)$ are known experimentally and presented in Table \ref{tab:results}. 
Combining these known factors with our lattice
result for $\text{Im} (A_2) / \text{Re}(A_2)$ we can determine the the unknown quantity Im($A_0$) within the Standard Model, finding 

\begin{equation}
\text{Im}(A_0) = -5.32(64)_{\text{stat}}(71)_{\text{syst}}\times 10^{-11}\text{ GeV}. 
\end{equation}
The error on Im$\,A_0$ is obtained by combining the errors on the quantities in Table \textref{tab:results}{\ref{tab:results}} in quadrature. 
In equation \eqref{eq:rearrange} below we compare the relative contribution to $\text{Im}(A_0)/\text{Re}(A_0)$ from $\text{Im}(A_2)/\text{Re}(A_2)$ 
and the term containing the experimentally known contributions:
\begin{equation}
\label{eq:rearrange}
 \begin{array}{ccccc}
  \dfrac{\text{Im}(A_0)}{\text{Re}(A_0)} & = &\dfrac{\text{Im}(A_2)}{\text{Re}(A_2)} &- &\dfrac{\sqrt{2} \left\vert \epsilon \right \vert}{\omega} \dfrac{\epsilon'}{\epsilon}\\
&&\\
-1.60(19)_{\mathrm{stat}}(21)_{\mathrm{syst}}\times10^{-4}  & = &-4.38(34)_{\mathrm{stat}}(95)_{\mathrm{syst}}\times 10^{-5} & -&1.16(18) \times 10^{-4}~.
 \end{array}
\end{equation}
Thus we see that while the error on the determination of Im$(A_0)$ is dominated by the uncertainty in the experimental value of $ \epsilon^\prime/ \epsilon$, 
the contribution of Im($A_2$)/Re($A_2$) to Im($A_0$) is significant (about 25\% in the determination of $\text{Im}(A_0)/\text{Re}(A_0)$). 

\begin{center}
\begin{table}[h]
\centering
\begin{tabular}[h]{|c|c|}
 \hline
$\text{Re}(\epsilon'/\epsilon)$ & $(1.65\pm0.26)\times10^{-3}$\\
$\omega$ & 0.04454(12)\\
$\left \vert \epsilon \right \vert$ & $(2.228\pm 0.011) \times 10^{-3}$ \\
Re$(A_0)$ & $3.3201(18)\times 10^{-7}$~GeV\\
$\text{Im} (A_2)/\text{Re}(A_2)$ (lattice) & $-4.38(34)_{\mathrm{stat}}(95)_{\mathrm{syst}} \times 10^{-5}$\\
\hline
\end{tabular}
\caption{\label{tab:results}  Experimental values of the components of equation (6.1) used in the determination of Im($A_0$), together with
the results for $\text{Im}(A_2)/\text{Re}(A_2)$ from these proceedings.}
\end{table}
\end{center}

\vspace{-0.06 in}
\section{Conclusions}
\vspace{-0.05 in}
We have presented preliminary results for the $\Delta I = 3/2$ $K \rightarrow \pi\pi$ decay amplitude on $32^3$ lattices with $2+1$ flavours
of DWF and the Iwasaki-DSDR gauge action. We find 
\begin{equation*}
\begin{split}
\text{Re}(A_2) & = (1.436\pm 0.063_{\text{stat}}\pm 0.258_{\text{syst}})\times 10^{-8}\text{ GeV},\\ 
\quad \text{Im}(A_2) &= (-6.29 \pm 0.46_{\text{stat}} \pm 1.20_{\text{syst}})\times 10^{-13}\text{ GeV}.
\end{split}
\end{equation*}
Our result for $\text{Re}(A_2)$ is in good 
agreement with the experimental result of $1.479(4) \times 10^{-8}\text{ GeV}$ obtained from $K^+$ decays. In the future we plan to undertake 
similar calculations of $\text{Re}(A_0)$ and $\text{Im}(A_0)$ \cite{qi:lattice2011}, allowing $\epsilon'/\epsilon$ to be calculated from first principles 
for the first time.
In the mean time we make the prediction, based on the Standard Model, that  
$\text{Im}(A_0) = -5.32(64)_{\text{stat}}(71)_{\text{sys}}\times 10^{-11}\text{ GeV}$.

We thank all of our colleagues in the RBC
and UKQCD collaborations for their contribution to these results, for
helpful discussions and the development and support of the QCDOC hardware and software infrastructure which was 
essential to this work. In addition we acknowledge Columbia University, RIKEN, BNL, ANL, and the U.S. DOE for providing the facilities on which 
this work was performed. This research used resources of the Argonne Leadership Computing Facility at Argonne National Laboratory, 
which is supported by the Office of Science of the U.S. DOE under contract DE-AC02-06CH11357. 
This work was supported in part by U.S. DOE grant number DE-FG02-92ER40699. E.G. is supported by an
STFC studentship and grant ST/G000557/1 and by EU contract MRTN-CT-2006-03542 (Flavianet).


\begin{thebibliography}{99}

\bibitem{Blum:2001xb}
  T.~Blum {\it et al.}  [RBC Collaboration],
  Phys.\ Rev.\  D {\bf 68} (2003) 114506
  [arXiv:hep-lat/0110075].

\bibitem{kim_christ}
C.H. Kim and N.H. Christ, \emph{Nucl. Phys. Proc. Suppl.} {\bf{119}} (2003) 365 [arXiv:hep-lat/0210003]
\bibitem{CPPACS}
J. Noaki et al., \emph{Phys. Rev. D} {\bf 68} (2003) 014501 [arXiv:hep-lat/0108013]
\bibitem{chiral_extrap}
N.H. Christ and Li, S., \pos{PoS(Lattice 2008)272},  [arXiv:hep-lat/08121368]

\bibitem{Goode:2010lat}
E. Goode and M. Lightman, \pos{PoS(Lattice 2010)313}, 2010
[arXiv:1101.2473v1 [hep-lat]]

\bibitem{LLfactor}
Laurent. Lellouch and Martin L\"uscher, \emph{Commun. Math. Phys} {\bf{219}} (2001) 31

\bibitem{LLfactor_sachrajda}
C.-J.D. Lin et al., \emph{Nucl. Phys. B}{\bf{619}} (2001) 467 [arXiv:hep-lat/0104006]

\bibitem{Luscher_quant}
M. L\"uscher, \emph{Nucl. Phys. B} {\bf{354, 531}} (1991) 

\bibitem{Schenk:1991xe}
  A.~Schenk,
  Nucl.\ Phys.\  B {\bf 363} (1991) 97.

\bibitem{Garron_PoS} 
N. Garron and A. T. Lytle, \pos{PoS(Lattice 2011)335}, 2011

\bibitem{Buchalla:1995vs}
  G.~Buchalla, A.~J.~Buras and M.~E.~Lautenbacher,
  Rev.\ Mod.\ Phys.\  {\bf 68} (1996) 1125
  [arXiv:hep-ph/9512380].

\bibitem{ktopipi}
T.~Blum {\it et al.}  [RBC Collaboration and UKQCD Collaboration],
manuscript in preparation

\bibitem{lightman_thesis}
M. Lightman, Doctoral Thesis (2011)
\bibitem{Aoki:2010dy}
  Y.~Aoki {\it et al.}  [RBC Collaboration and UKQCD Collaboration],
  Phys.\ Rev.\  D {\bf 83} (2011) 074508
  [arXiv:1011.0892 [hep-lat]].

\bibitem{qi:lattice2011}
Q. Liu, \pos{PoS(Lattice 2011)287}, [arXiv:1110.2143 [hep-lat]]
\end{thebibliography}
\end{document}